%
%
%
\def\unredoffs{} \def\redoffs{\voffset=-.31truein\hoffset=-.48truein}
\def\speclscape{}
%
%
%
%
%
\newbox\leftpage \newdimen\fullhsize \newdimen\hstitle \newdimen\hsbody
\tolerance=1000\hfuzz=2pt
\catcode`\@=11 
\ifx\hyperdef\UNd@FiNeD\def\hyperdef#1#2#3#4{#4}\def\hyperref#1#2#3#4{#4}\fi
\def\bigans{b }
\def\answ{b }
%
\ifx\answ\bigans\message{(This will come out unreduced.}
\magnification=1200\unredoffs\baselineskip=16pt plus 2pt minus 1pt
\hsbody=\hsize \hstitle=\hsize 
\else\message{(This will be reduced.} \let\l@r=L
\magnification=1000\baselineskip=16pt plus 2pt minus 1pt \vsize=7truein
\redoffs \hstitle=8truein\hsbody=4.75truein\fullhsize=10truein\hsize=\hsbody
\output={\ifnum\pageno=0 
  \shipout\vbox{\speclscape{\hsize\fullhsize\makeheadline}
    \hbox to \fullhsize{\hfill\pagebody\hfill}}\advancepageno
  \else
  \almostshipout{\leftline{\vbox{\pagebody\makefootline}}}\advancepageno
  \fi}
\def\almostshipout#1{\if L\l@r \count1=1 \message{[\the\count0.\the\count1]}
      \global\setbox\leftpage=#1 \global\let\l@r=R
 \else \count1=2
  \shipout\vbox{\speclscape{\hsize\fullhsize\makeheadline}
      \hbox to\fullhsize{\box\leftpage\hfil#1}}  \global\let\l@r=L\fi}
\fi
%
\newcount\yearltd\yearltd=\year\advance\yearltd by -2000

\def\Title#1#2{\nopagenumbers\abstractfont\hsize=\hstitle\rightline{#1}%
\vskip 1in\centerline{\titlefont #2}\abstractfont\vskip .5in\pageno=0}
\def\Date#1{\vfill\leftline{#1}\tenpoint\supereject\global\hsize=\hsbody%
\footline={\hss\tenrm\hyperdef\hypernoname{page}\folio\folio\hss}}%
%

\def\draftmode{\message{ DRAFTMODE }\def\draftdate{{\rm preliminary draft:
\number\month/\number\day/\number\yearltd\ \ \hourmin}}%
\headline={\hfil\draftdate}\writelabels\baselineskip=20pt plus 2pt minus 2pt
 {\count255=\time\divide\count255 by 60 \xdef\hourmin{\number\count255}
  \multiply\count255 by-60\advance\count255 by\time
  \xdef\hourmin{\hourmin:\ifnum\count255<10 0\fi\the\count255}}}
\def\nolabels{\def\wrlabeL##1{}\def\eqlabeL##1{}\def\reflabeL##1{}}
\def\writelabels{\def\wrlabeL##1{\leavevmode\vadjust{\rlap{\smash%
{\line{{\escapechar=` \hfill\rlap{\sevenrm\hskip.03in\string##1}}}}}}}%
\def\eqlabeL##1{{\escapechar-1\rlap{\sevenrm\hskip.05in\string##1}}}%
\def\reflabeL##1{\noexpand\llap{\noexpand\sevenrm\string\string\string##1}}}
\nolabels
%
\global\newcount\secno \global\secno=0
\global\newcount\meqno \global\meqno=1
\def\s@csym{}
\def\newsec#1{\global\advance\secno by1%
{\toks0{#1}\message{(\the\secno. \the\toks0)}}%
\global\subsecno=0\eqnres@t\let\s@csym\secsym\xdef\secn@m{\the\secno}\noindent
{\bf\hyperdef\hypernoname{section}{\the\secno}{\the\secno.} #1}%
\writetoca{{\string\hyperref{}{section}{\the\secno}{\it\the\secno.}} {{\it #1} }}%
\par\nobreak\medskip\nobreak}
\def\eqnres@t{\xdef\secsym{\the\secno.}\global\meqno=1\bigbreak\bigskip}
\def\sequentialequations{\def\eqnres@t{\bigbreak}}\xdef\secsym{}
\global\newcount\subsecno \global\subsecno=0
\def\subsec#1{\global\advance\subsecno by1%
{\toks0{#1}\message{(\s@csym\the\subsecno. \the\toks0)}}%
\ifnum\lastpenalty>9000\else\bigbreak\fi       \global\subsubsecno=0
\noindent{\it\hyperdef\hypernoname{subsection}{\secn@m.\the\subsecno}%
{\secn@m.\the\subsecno.} #1}\writetoca{\string\quad
{\string\hyperref{}{subsection}{\secn@m.\the\subsecno}{\secn@m.\the\subsecno.}}
{#1}}\par\nobreak\medskip\nobreak}
\def\appendix#1#2{\global\meqno=1\global\subsecno=0\xdef\secsym{\hbox{#1.}}%
\bigbreak\bigskip\noindent{\bf Appendix \hyperdef\hypernoname{appendix}{#1}%
{#1.} #2}{\toks0{(#1. #2)}\message{\the\toks0}}%
\xdef\s@csym{#1.}\xdef\secn@m{#1}%
\writetoca{\string\hyperref{}{appendix}{#1}{{\it Appendix} {\it #1.}} {\it #2}}%
\par\nobreak\medskip\nobreak}
%
%
\def\checkm@de#1#2{\ifmmode{\def\f@rst##1{##1}\hyperdef\hypernoname{equation}%
{#1}{#2}}\else\hyperref{}{equation}{#1}{#2}\fi}
\def\eqnn#1{\DefWarn#1\xdef #1{(\noexpand\relax\noexpand\checkm@de%
{\s@csym\the\meqno}{\secsym\the\meqno})}%
\wrlabeL#1\writedef{#1\leftbracket#1}\global\advance\meqno by1}
\def\f@rst#1{\c@t#1a\em@ark}\def\c@t#1#2\em@ark{#1}
\def\eqna#1{\DefWarn#1\wrlabeL{#1$\{\}$}%
\xdef #1##1{(\noexpand\relax\noexpand\checkm@de%
{\s@csym\the\meqno\noexpand\f@rst{##1}}{\hbox{$\secsym\the\meqno##1$}})}
\writedef{#1\numbersign1\leftbracket#1{\numbersign1}}\global\advance\meqno by1}
\def\eqn#1#2{\DefWarn#1%
\xdef #1{(\noexpand\hyperref{}{equation}{\s@csym\the\meqno}%
{\secsym\the\meqno})}$$#2\eqno(\hyperdef\hypernoname{equation}%
{\s@csym\the\meqno}{\secsym\the\meqno})\eqlabeL#1$$%
\writedef{#1\leftbracket#1}\global\advance\meqno by1}
\def\xeqn{\expandafter\xe@n}\def\xe@n(#1){#1}
\def\xeqna#1{\expandafter\xe@n#1}
\def\eqns#1{(\e@ns #1{\hbox{}})}
\def\e@ns#1{\ifx\UNd@FiNeD#1\message{eqnlabel \string#1 is undefined.}%
\xdef#1{(?.?)}\fi{\let\hyperref=\relax\xdef\next{#1}}%
\ifx\next\em@rk\def\next{}\else%
\ifx\next#1\xeqn#1\else\def\n@xt{#1}\ifx\n@xt\next#1\else\xeqna#1\fi
\fi\let\next=\e@ns\fi\next}

\def\DefWarn#1{\ifx\UNd@FiNeD#1\else
\immediate\write16{*** WARNING: the label \string#1 is already defined ***}\fi}
%
\newskip\footskip\footskip14pt plus 1pt minus 1pt 
\def\footnotefont{\ninepoint}\def\f@t#1{\footnotefont #1\@foot}
\def\f@@t{\baselineskip\footskip\bgroup\footnotefont\aftergroup\@foot\let\next}
\setbox\strutbox=\hbox{\vrule height9.5pt depth4.5pt width0pt}
\global\newcount\ftno \global\ftno=0
\def\foot{\global\advance\ftno by1\def\foot@rg{\hyperref{}{footnote}%
{\the\ftno}{\the\ftno}\xdef\foot@rg{\noexpand\hyperdef\noexpand\hypernoname%
{footnote}{\the\ftno}{\the\ftno}}}\footnote{$^{\foot@rg}$}}
%
\newwrite\ftfile
\def\footend{\def\foot{\global\advance\ftno by1\chardef\wfile=\ftfile
\hyperref{}{footnote}{\the\ftno}{$^{\the\ftno}$}%
\ifnum\ftno=1\immediate\openout\ftfile=\jobname.fts\fi%
\immediate\write\ftfile{\noexpand\smallskip%
\noexpand\item{\noexpand\hyperdef\noexpand\hypernoname{footnote}
{\the\ftno}{f\the\ftno}:\ }\pctsign}\findarg}%
\def\footatend{\vfill\eject\immediate\closeout\ftfile{\parindent=20pt
\centerline{\bf Footnotes}\nobreak\bigskip\input \jobname.fts }}}
\def\footatend{}
%
%
\global\newcount\refno \global\refno=1
\newwrite\rfile
\def\ref{[\hyperref{}{reference}{\the\refno}{\the\refno}]\nref}
\def\nref#1{\DefWarn#1%
\xdef#1{[\noexpand\hyperref{}{reference}{\the\refno}{\the\refno}]}%
\writedef{#1\leftbracket#1}%
\ifnum\refno=1\immediate\openout\rfile=\jobname.refs\fi
\chardef\wfile=\rfile\immediate\write\rfile{\noexpand\item{[\noexpand\hyperdef%
\noexpand\hypernoname{reference}{\the\refno}{\the\refno}]\ }%
\reflabeL{#1\hskip.31in}\pctsign}\global\advance\refno by1\findarg}
\def\findarg#1#{\begingroup\obeylines\newlinechar=`\^^M\pass@rg}
{\obeylines\gdef\pass@rg#1{\writ@line\relax #1^^M\hbox{}^^M}%
\gdef\writ@line#1^^M{\expandafter\toks0\expandafter{\striprel@x #1}%
\edef\next{\the\toks0}\ifx\next\em@rk\let\next=\endgroup\else\ifx\next\empty%
\else\immediate\write\wfile{\the\toks0}\fi\let\next=\writ@line\fi\next\relax}}
\def\striprel@x#1{} \def\em@rk{\hbox{}}
\def\lref{\begingroup\obeylines\lr@f}
\def\lr@f#1#2{\DefWarn#1\gdef#1{\let#1=\UNd@FiNeD\ref#1{#2}}\endgroup\unskip}

\def\addref#1{\immediate\write\rfile{\noexpand\item{}#1}} 
\def\listrefs{\footatend\vfill\supereject\immediate\closeout\rfile\writestoppt
\baselineskip=\footskip\centerline{{\bf References}}\bigskip{\parindent=20pt%
\frenchspacing\escapechar=` \input \jobname.refs\vfill\eject}\nonfrenchspacing}
\def\startrefs#1{\immediate\openout\rfile=\jobname.refs\refno=#1}
\def\xref{\expandafter\xr@f}\def\xr@f[#1]{#1}
\def\refs#1{\count255=1[\r@fs #1{\hbox{}}]}
\def\r@fs#1{\ifx\UNd@FiNeD#1\message{reflabel \string#1 is undefined.}%
\nref#1{need to supply reference \string#1.}\fi%
\vphantom{\hphantom{#1}}{\let\hyperref=\relax\xdef\next{#1}}%
\ifx\next\em@rk\def\next{}%
\else\ifx\next#1\ifodd\count255\relax\xref#1\count255=0\fi%
\else#1\count255=1\fi\let\next=\r@fs\fi\next}
%

%
\newwrite\ffile\global\newcount\figno \global\figno=1
\def\fig{fig.~\hyperref{}{figure}{\the\figno}{\the\figno}\nfig}
\def\nfig#1{\DefWarn#1%
\xdef#1{fig.~\noexpand\hyperref{}{figure}{\the\figno}{\the\figno}}%
\writedef{#1\leftbracket fig.\noexpand~\xfig#1}%
\ifnum\figno=1\immediate\openout\ffile=\jobname.figs\fi\chardef\wfile=\ffile%
{\let\hyperref=\relax
\immediate\write\ffile{\noexpand\medskip\noexpand\item{Fig.\ %
\noexpand\hyperdef\noexpand\hypernoname{figure}{\the\figno}{\the\figno}. }
\reflabeL{#1\hskip.55in}\pctsign}}\global\advance\figno by1\findarg}
\def\listfigs{\vfill\eject\immediate\closeout\ffile{\parindent40pt
\baselineskip14pt\centerline{{\bf Figure Captions}}\nobreak\medskip
\escapechar=` \input \jobname.figs\vfill\eject}}
\def\xfig{\expandafter\xf@g}\def\xf@g fig.\penalty\@M\ {}
\def\figs#1{figs.~\f@gs #1{\hbox{}}}
\def\f@gs#1{{\let\hyperref=\relax\xdef\next{#1}}\ifx\next\em@rk\def\next{}\else
\ifx\next#1\xfig #1\else#1\fi\let\next=\f@gs\fi\next}
\def\figin{\epsfcheck\figin}\def\figins{\epsfcheck\figins}
\def\epsfcheck{\ifx\epsfbox\UNd@FiNeD
\message{(NO epsf.tex, FIGURES WILL BE IGNORED)}
\gdef\figin##1{\vskip2in}\gdef\figins##1{\hskip.5in}
\else\message{(FIGURES WILL BE INCLUDED)}%
\gdef\figin##1{##1}\gdef\figins##1{##1}\fi}
\def\DefWarn#1{}
\def\figinsert{\goodbreak\midinsert}
\def\ifig#1#2#3{\DefWarn#1\xdef#1{Fig.~\noexpand\hyperref{}{figure}%
{\the\figno}{\the\figno}}\writedef{#1\leftbracket fig.\noexpand~\xfig#1}%
\figinsert\figin{\centerline{#3}}\medskip\centerline{\vbox{\baselineskip12pt
\advance\hsize by -1truein\noindent\wrlabeL{#1=#1}\footnotefont%
{\bf Fig.~\hyperdef\hypernoname{figure}{\the\figno}{\the\figno}:} #2}}
\bigskip\endinsert\global\advance\figno by1}
\newwrite\lfile
{\escapechar-1\xdef\pctsign{\string\%}\xdef\leftbracket{\string\{}
\xdef\rightbracket{\string\}}\xdef\numbersign{\string\#}}
\def\writedefs{\immediate\openout\lfile=\jobname.defs \def\writedef##1{%
{\let\hyperref=\relax\let\hyperdef=\relax\let\hypernoname=\relax
 \immediate\write\lfile{\string\def\string##1\rightbracket}}}}%
\def\writestop{\def\writestoppt{\immediate\write\lfile{\string\pageno
 \the\pageno\string\startrefs\leftbracket\the\refno\rightbracket
 \string\def\string\secsym\leftbracket\secsym\rightbracket
 \string\secno\the\secno\string\meqno\the\meqno}\immediate\closeout\lfile}}
\def\writestoppt{}\def\writedef#1{}
\def\seclab#1{\DefWarn#1%
\xdef #1{\noexpand\hyperref{}{section}{\the\secno}{\the\secno}}%
\writedef{#1\leftbracket#1}\wrlabeL{#1=#1}}
\def\subseclab#1{\DefWarn#1%
\xdef #1{\noexpand\hyperref{}{subsection}{\secn@m.\the\subsecno}%
{\secn@m.\the\subsecno}}\writedef{#1\leftbracket#1}\wrlabeL{#1=#1}}
\def\applab#1{\DefWarn#1%
\xdef #1{\noexpand\hyperref{}{appendix}{\secn@m}{\secn@m}}%
\writedef{#1\leftbracket#1}\wrlabeL{#1=#1}}
\newwrite\tfile \def\writetoca#1{}
\def\leaderfill{\leaders\hbox to 1em{\hss.\hss}\hfill}
\def\writetoc{\immediate\openout\tfile=\jobname.toc
   \def\writetoca##1{{\edef\next{\write\tfile{\noindent ##1
   \string\leaderfill {\string\hyperref{}{page}{\noexpand\number\pageno}%
                       {\noexpand\number\pageno}} \par}}\next}}}
\newread\ch@ckfile
\def\listtoc{\immediate\closeout\tfile\immediate\openin\ch@ckfile=\jobname.toc
\ifeof\ch@ckfile\message{no file \jobname.toc, no table of contents this pass}%
\else\closein\ch@ckfile\centerline{\bf Contents}\nobreak\medskip%
{\baselineskip=18.5pt  \footnotefont
\parskip=2pt\catcode`\@=12\input\jobname.toc
\catcode`\@=12\bigbreak\bigskip}\fi}
\catcode`\@=12 
%
\edef\tfontsize{\ifx\answ\bigans scaled\magstep3\else scaled\magstep4\fi}
\font\titlerm=cmr10 \tfontsize \font\titlerms=cmr7 \tfontsize
\font\titlermss=cmr5 \tfontsize \font\titlei=cmmi10 \tfontsize
\font\titleis=cmmi7 \tfontsize \font\titleiss=cmmi5 \tfontsize
\font\titlesy=cmsy10 \tfontsize \font\titlesys=cmsy7 \tfontsize
\font\titlesyss=cmsy5 \tfontsize \font\titleit=cmti10 \tfontsize
\skewchar\titlei='177 \skewchar\titleis='177 \skewchar\titleiss='177
\skewchar\titlesy='60 \skewchar\titlesys='60 \skewchar\titlesyss='60
\def\titlefont{\def\rm{\fam0\titlerm}
\textfont0=\titlerm \scriptfont0=\titlerms \scriptscriptfont0=\titlermss
\textfont1=\titlei \scriptfont1=\titleis \scriptscriptfont1=\titleiss
\textfont2=\titlesy \scriptfont2=\titlesys \scriptscriptfont2=\titlesyss
\textfont\itfam=\titleit \def\it{\fam\itfam\titleit}\rm}
 \ifx\answ\bigans\else scaled\magstep1\fi
\ifx\answ\bigans\def\abstractfont{\tenpoint}\else
\font\absit=cmti10 scaled \magstep1
\font\abssl=cmsl10 scaled \magstep1
\font\absrm=cmr10 scaled\magstep1 \font\absrms=cmr7 scaled\magstep1
\font\absrmss=cmr5 scaled\magstep1 \font\absi=cmmi10 scaled\magstep1
\font\absis=cmmi7 scaled\magstep1 \font\absiss=cmmi5 scaled\magstep1
\font\abssy=cmsy10 scaled\magstep1 \font\abssys=cmsy7 scaled\magstep1
\font\abssyss=cmsy5 scaled\magstep1 \font\absbf=cmbx10 scaled\magstep1
\skewchar\absi='177 \skewchar\absis='177 \skewchar\absiss='177
\skewchar\abssy='60 \skewchar\abssys='60 \skewchar\abssyss='60
\def\abstractfont{\def\rm{\fam0\absrm}
\textfont0=\absrm \scriptfont0=\absrms \scriptscriptfont0=\absrmss
\textfont1=\absi \scriptfont1=\absis \scriptscriptfont1=\absiss
\textfont2=\abssy \scriptfont2=\abssys \scriptscriptfont2=\abssyss
\textfont\itfam=\absit \def\it{\fam\itfam\absit}\def\footnotefont{\tenpoint}%
\textfont\slfam=\abssl \def\sl{\fam\slfam\abssl}%
\textfont\bffam=\absbf \def\bf{\fam\bffam\absbf}\rm}\fi
\def\tenpoint{\def\rm{\fam0\tenrm}
\textfont0=\tenrm \scriptfont0=\sevenrm \scriptscriptfont0=\fiverm
\textfont1=\teni  \scriptfont1=\seveni  \scriptscriptfont1=\fivei
\textfont2=\tensy \scriptfont2=\sevensy \scriptscriptfont2=\fivesy
\textfont\itfam=\tenit \def\it{\fam\itfam\tenit}\def\footnotefont{\ninepoint}%
\textfont\bffam=\tenbf \def\bf{\fam\bffam\tenbf}\def\sl{\fam\slfam\tensl}\rm}
\font\ninerm=cmr9 \font\sixrm=cmr6 \font\ninei=cmmi9 \font\sixi=cmmi6
\font\ninesy=cmsy9 \font\sixsy=cmsy6 \font\ninebf=cmbx9
\font\nineit=cmti9 \font\ninesl=cmsl9 \skewchar\ninei='177
\skewchar\sixi='177 \skewchar\ninesy='60 \skewchar\sixsy='60
\def\ninepoint{\def\rm{\fam0\ninerm}
\textfont0=\ninerm \scriptfont0=\sixrm \scriptscriptfont0=\fiverm
\textfont1=\ninei \scriptfont1=\sixi \scriptscriptfont1=\fivei
\textfont2=\ninesy \scriptfont2=\sixsy \scriptscriptfont2=\fivesy
\textfont\itfam=\ninei \def\it{\fam\itfam\nineit}\def\sl{\fam\slfam\ninesl}%
\textfont\bffam=\ninebf \def\bf{\fam\bffam\ninebf}\rm}
%
%
\def\noblackbox{\overfullrule=0pt}
\hyphenation{anom-aly anom-alies coun-ter-term coun-ter-terms}
\def\inv{^{\raise.15ex\hbox{${\scriptscriptstyle -}$}\kern-.05em 1}}

\def\Dsl{\,\raise.15ex\hbox{/}\mkern-13.5mu D} 
\def\dsl{\raise.15ex\hbox{/}\kern-.57em\partial}

\def\lspace{\ifx\answ\bigans{}\else\qquad\fi}
\def\lbspace{\ifx\answ\bigans{}\else\hskip-.2in\fi} 
\def\boxeqn#1{\vcenter{\vbox{\hrule\hbox{\vrule\kern3pt\vbox{\kern3pt
	\hbox{${\displaystyle #1}$}\kern3pt}\kern3pt\vrule}\hrule}}}
\def\mbox#1#2{\vcenter{\hrule \hbox{\vrule height#2in
		\kern#1in \vrule} \hrule}}  
%

\def\darr#1{\raise1.5ex\hbox{$\leftrightarrow$}\mkern-16.5mu #1}

\def\roughly#1{\raise.3ex\hbox{$#1$\kern-.75em\lower1ex\hbox{$\sim$}}}

\global\newcount\subsubsecno \global\subsubsecno=0
\def\subsubsec#1{\global\advance\subsubsecno by1%
{\toks0{#1}\message{(\the\secno\the\subsecno\the\subsubsecno. \the\toks0)}}%
\ifnum\lastpenalty>9000\else\bigbreak\fi
\noindent{\it\hyperdef\hypernoname{subsubsection}{\the\secno.\the\subsecno\the\subsubsecno}%
{\the\secno.\the\subsecno.\the\subsubsecno.} #1}
\par\nobreak\medskip\nobreak}
\def\boxit#1{\vbox{\hrule\hbox{\vrule\kern8pt
\vbox{\hbox{\kern8pt}\hbox{\vbox{#1}}\hbox{\kern8pt}}
\kern8pt\vrule}\hrule}}
\def\mathboxit#1{\vbox{\hrule\hbox{\vrule\kern8pt\vbox{\kern8pt
\hbox{$\displaystyle #1$}\kern8pt}\kern8pt\vrule}\hrule}}
\def\slashchar#1{\setbox0=\hbox{$#1$}           
   \dimen0=\wd0                                 
   \setbox1=\hbox{/} \dimen1=\wd1               
   \ifdim\dimen0>\dimen1                        
      \rlap{\hbox to \dimen0{\hfil/\hfil}}      
      #1                                        
   \else                                        
      \rlap{\hbox to \dimen1{\hfil$#1$\hfil}}   
      /                                         
   \fi}
\def\sqr#1#2{{\vcenter{\vbox{\hrule height.#2pt
         \hbox{\vrule width.#2pt height#1pt \kern#1pt
            \vrule width.#2pt}
         \hrule height.#2pt}}}}


\input amssym.def
\input amssym.tex
\noblackbox
\baselineskip=14.5pt
\def\crr{\noalign{\vskip5pt}}

\def\comment#1{{}}

\def\ap{\alpha'}

\def\ie{{ i.e.\ }}

\def\eqq{{\it Eq.\ }}

\def\al{\alpha}

\def\Om{\Omega}
\def\bet{\beta}

\newif\ifnref

\nreffalse

\input epsf

\def\figin{\epsfcheck\figin}\def\figins{\epsfcheck\figins}
\def\epsfcheck{\ifx\epsfbox\UnDeFiNeD
\message{(NO epsf.tex, FIGURES WILL BE IGNORED)}
\gdef\figin##1{\vskip2in}\gdef\figins##1{\hskip.5in}
\else\message{(FIGURES WILL BE INCLUDED)}%
\gdef\figin##1{##1}\gdef\figins##1{##1}\fi}
\def\DefWarn#1{}
\def\figinsert{\goodbreak\midinsert}  
\def\ifig#1#2#3{\DefWarn#1\xdef#1{Fig.~\the\figno}
\writedef{#1\leftbracket fig.\noexpand~\the\figno}%
\figinsert\figin{\centerline{#3}}\medskip\centerline{\vbox{\baselineskip12pt
\advance\hsize by -1truein\noindent\footnotefont\centerline{{\bf
Fig.~\the\figno}\ \sl #2}}}
\bigskip\endinsert\global\advance\figno by1}

\def\iifig#1#2#3#4{\DefWarn#1\xdef#1{Fig.~\the\figno}
\writedef{#1\leftbracket fig.\noexpand~\the\figno}%
\figinsert\figin{\centerline{#4}}\medskip\centerline{\vbox{\baselineskip12pt
\advance\hsize by -1truein\noindent\footnotefont\centerline{{\bf
Fig.~\the\figno}\ \ \sl #2}}}\smallskip\centerline{\vbox{\baselineskip12pt
\advance\hsize by -1truein\noindent\footnotefont\centerline{\ \ \ \sl #3}}}
\bigskip\endinsert\global\advance\figno by1}


\def\tilde{\widetilde}

\def\h {{1\over 2}}

\def\ov {\overline}
\def\o {\over}
\def\fc#1#2{{#1 \o #2}}

\def\IC{{\bf C}}\def\IR{ {\bf R}}


\def\br{\hfill\break}

\def\det {{\rm det}}

\def\lf {\left}
\def\ri {\right}
\def\ra {\rightarrow}

\def\re {{\rm Re}}
\def\im {{\rm Im}}
\def\p {\partial}

 \def\Sc {{\cal S}}

\def\ceiling#1{\lceil#1\rceil}
\def\floor#1{\lfloor#1\rfloor}



\def\IH{{\bf H}_+}

\lref\StiebergerWEA{
  S.~Stieberger,
``Closed superstring amplitudes, single-valued multiple zeta values and the Deligne associator,''
J.\ Phys.\ A {\bf 47}, 155401 (2014).
[arXiv:1310.3259 [hep-th]].
}

\lref\StiebergerHBA{
  S.~Stieberger and T.R.~Taylor,
 ``Closed String Amplitudes as Single-Valued Open String Amplitudes,''
Nucl.\ Phys.\ B {\bf 881}, 269 (2014).
[arXiv:1401.1218 [hep-th]].
}

\lref\KleissNE{
  R.~Kleiss and H.~Kuijf,
  ``Multi - Gluon Cross-sections and Five Jet Production at Hadron Colliders,''
Nucl.\ Phys.\ B {\bf 312}, 616 (1989)..
}

\lref\StiebergerHQ{
  S.~Stieberger,
``Open \& Closed vs. Pure Open String Disk Amplitudes,''
[arXiv:0907.2211 [hep-th]].
}

\lref\stnew{
  S.~Stieberger and T.R.~Taylor,
  ``Disk Scattering of Open and Closed Strings (II),''
  in preparation.}

\lref\KawaiXQ{
  H.~Kawai, D.C.~Lewellen and S.H.H.~Tye,
``A Relation Between Tree Amplitudes Of Closed And Open Strings,''
  Nucl.\ Phys.\  B {\bf 269}, 1 (1986).
}

\lref\StiebergerKIA{
  S.~Stieberger and T.R.~Taylor,
``Subleading Terms in the Collinear Limit of Yang-Mills Amplitudes,''
[arXiv:1508.01116 [hep-th]], to appear in Phys.\ Lett.\ B.
}

\lref\StiebergerTE{
  S.~Stieberger and T.R.~Taylor,
``Multi-Gluon Scattering in Open Superstring Theory,''
Phys.\ Rev.\ D {\bf 74}, 126007 (2006).
[hep-th/0609175].
}

\lref\MafraNVii{
C.R.~Mafra, O.~Schlotterer and S.~Stieberger,
``Complete N-Point Superstring Disk Amplitude II. Amplitude and Hypergeometric Function Structure,''
Nucl.\ Phys.\ B {\bf 873}, 461 (2013).
[arXiv:1106.2646 [hep-th]].
}

\lref\BernQJ{
  Z.~Bern, J.J.~M.~Carrasco and H.~Johansson,
  ``New Relations for Gauge-Theory Amplitudes,''
Phys.\ Rev.\ D {\bf 78}, 085011 (2008).
[arXiv:0805.3993 [hep-ph]].
}

\lref\notation{M.L.~Mangano and S.J.~Parke,
``Multiparton amplitudes in gauge theories,''
Phys. Rept.  {\bf 200}, 301 (1991).
[hep-th/0509223];\br
L.J.~Dixon,
  ``Calculating scattering amplitudes efficiently,''
in Boulder 1995, QCD and beyond 539-582.
[hep-ph/9601359].}

\lref\KosteleckyPX{
  V.A.~Kostelecky, O.~Lechtenfeld and S.~Samuel,
``Covariant String Amplitudes On Exotic Topologies To One Loop,''
Nucl.\ Phys.\ B {\bf 298}, 133 (1988).
}

\lref\StiebergerTE{
  S.~Stieberger and T.R.~Taylor,
``Multi-Gluon Scattering in Open Superstring Theory,''
Phys.\ Rev.\ D {\bf 74}, 126007 (2006).
[hep-th/0609175].
}

\lref\BjerrumBohrRD{
  N.E.J.~Bjerrum-Bohr, P.H.~Damgaard and P.~Vanhove,
``Minimal Basis for Gauge Theory Amplitudes,''
Phys.\ Rev.\ Lett.\  {\bf 103}, 161602 (2009).
[arXiv:0907.1425 [hep-th]].
}

\lref\Bohr{
  N.E.J.~Bjerrum-Bohr, P.H.~Damgaard, T.~Sondergaard and P.~Vanhove,
 ``The Momentum Kernel of Gauge and Gravity Theories,''
JHEP {\bf 1101}, 001 (2011).
[arXiv:1010.3933 [hep-th]].
}

\lref\BernSV{
  Z.~Bern, L.J.~Dixon, M.~Perelstein and J.S.~Rozowsky,
``Multileg one loop gravity amplitudes from gauge theory,''
Nucl.\ Phys.\ B {\bf 546}, 423 (1999).
[hep-th/9811140].
}

\lref\StiebergerCEA{
  S.~Stieberger and T.R.~Taylor,
``Graviton as a Pair of Collinear Gauge Bosons,''
Phys.\ Lett.\ B {\bf 739}, 457 (2014).
[arXiv:1409.4771 [hep-th]].
}

\lref\StiebergerQJA{
  S. Stieberger and T.R.~Taylor,
``Graviton Amplitudes from Collinear Limits of Gauge Amplitudes,''
Phys.\ Lett.\ B {\bf 744}, 160 (2015).
[arXiv:1502.00655 [hep-th]].
}

\lref\HashimotoKF{
  A.~Hashimoto and I.~R.~Klebanov,
``Decay of excited D-branes,''
Phys.\ Lett.\ B {\bf 381}, 437 (1996).
[hep-th/9604065].
}

\lref\GubserWT{
  S.S.~Gubser, A.~Hashimoto, I.R.~Klebanov and J.M.~Maldacena,
``Gravitational lensing by $p$-branes,''
Nucl.\ Phys.\ B {\bf 472}, 231 (1996).
[hep-th/9601057].
}

\lref\GarousiAD{
  M.R.~Garousi and R.C.~Myers,
``Superstring scattering from D-branes,''
Nucl.\ Phys.\ B {\bf 475}, 193 (1996).
[hep-th/9603194].
}

\lref\KlebanovNI{
  I.R.~Klebanov and L.~Thorlacius,
``The Size of p-branes,''
Phys.\ Lett.\ B {\bf 371}, 51 (1996).
[hep-th/9510200].
}

\lref\GarousiEA{
  M.R.~Garousi and R.C.~Myers,
``World volume potentials on D-branes,''
JHEP {\bf 0011}, 032 (2000).
[hep-th/0010122].
}

\lref\LustCX{
  D.~L\"ust, P.~Mayr, R.~Richter and S.~Stieberger,
``Scattering of gauge, matter, and moduli fields from intersecting branes,''
Nucl.\ Phys.\ B {\bf 696}, 205 (2004).
[hep-th/0404134].
}

\Title{\vbox{\rightline{MPP--2015--184}
}}
{\vbox{\centerline{Disk Scattering of Open and Closed Strings (I)}}}
\medskip
\centerline{Stephan Stieberger$^a$ and Tomasz R. Taylor$^b$}
\bigskip
\centerline{\it $^a$ Max--Planck--Institut f\"ur Physik}
\centerline{\it Werner--Heisenberg--Institut, 80805 M\"unchen, Germany}
\medskip
\centerline{\it  $^b$ Department of Physics}
\centerline{\it  Northeastern University, Boston, MA 02115, USA}

\vskip15pt

\medskip
\bigskip\bigskip\bigskip
\centerline{\bf Abstract}
\vskip .2in
\noindent

\noindent
At the tree level, the scattering processes involving open and closed strings are described by a disk world--sheet
with vertex operator insertions at the boundary and in the bulk.
Such amplitudes can be decomposed as certain linear combinations of pure open string amplitudes.
While previous relations have been established on the double cover (complex sphere) in this letter we derive them  on the disk (upper complex half plane) allowing for different momenta of the
left-- and right--movers of the closed string.
Formally, the computation of disk amplitudes involving both open and closed strings is reduced
to considering the monodromies on the underlying string world--sheet.

\Date{}
\noindent
\goodbreak
\break

The relationship between open and closed string amplitudes is important from both mathematical and physical points of view because it helps in understanding what
features of the closed string can be implemented by pure open string properties.
At tree--level, Kawai, Lewellen and Tye (KLT) \KawaiXQ\ derived a formula which expresses any closed string tree amplitude in terms of a sum of the products of appropriate open string tree amplitudes.
This formula gives rise to a striking relation between gravity and gauge amplitudes at tree--level.
An other description has been developed in
\refs{\StiebergerWEA,\StiebergerHBA}, by constructing tree--level closed superstring amplitudes through the  ``single--valued'' projection of open superstring amplitudes. This projection yields linear relations between the functions encompassing effects of massive closed and open superstring excitations, to all orders in the inverse string tension $\ap$.
They reveal a deeper connection between gauge and gravity string amplitudes than what is implied by the KLT relations.
Furthermore, in \StiebergerHQ\ tree--level string amplitudes involving both open and closed strings have  been expressed as linear combinations of pure open string amplitudes.
This correspondence gives a  relation between Einstein--Yang--Mills (EYM) theory and pure gauge amplitudes at tree--level \StiebergerCEA\ with interesting consequences
for constructing  gravity amplitudes from gauge amplitudes \StiebergerQJA.
Scattering amplitudes of open and closed strings describe the couplings of brane and bulk fields thus probing the effective  D--brane action. Hence, these amplitudes are important for many studies related to D--brane effects.

Tree--level amplitudes involving both open and closed strings are described by a disk world--sheet,
which is an oriented manifold with one boundary.
The latter can be mapped to the upper half plane:
\eqn\upper{
\IH=\{z\in \IC\ |\ \im(z)\geq 0\ \}\ .}
Open string vertex operator insertions are placed at the boundary of the disk and closed string
positions at the bulk. The integration over the latter can be extended from the half--plane covering the disk to the full complex plane if the closed strings are world--sheet symmetric closed string states (such as graviton or dilaton). However, for arbitrary closed string states and generic
D--brane and orientifold configurations  this world--sheet symmetry is not furnished
and one has to perform the computations  on the disk.
The techniques for evaluating generic disk integrals involving both open and (world--sheet symmetric) closed string states have been developed in \StiebergerHQ.
Moreover, in \StiebergerCEA\ a closed and compact expression for
the amplitude involving one closed and any number of open strings has been derived. In this letter
we want to extend these results to generic closed string states, i.e. perform the amplitude computation on the disk rather than on its double cover.
The amplitudes can be decomposed as certain linear combinations of pure open string amplitudes.
Formally, the computation of disk amplitudes involving both open and closed strings is reduced
to considering the monodromies on the underlying string world--sheet.

In the following we shall consider disk amplitudes with one bulk and $N-2$ boundary operators\foot{Disk amplitudes with an arbitrary number of  bulk and  boundary operators will be considered in~\stnew.}.
This yields the leading order amplitude for either the absorption of a closed string by a D--brane
or the decay of an excited D--brane into a massless closed string state and the unexcited D--brane
\refs{\KlebanovNI,\GubserWT,\HashimotoKF}.
Open string vertices  with momenta $p_i,~i=1,\dots,N{-}2$ are inserted on the real axis of \upper\ at
$x_i\in\IR$,  while a single closed string vertex operator  is inserted at complex $z\in\IH$.
For the latter we assume different left-- and right--moving space--time momenta $q_1$ and $q_2$, respectively. This is the most general setup for scattering both open and closed strings in the presence of D-branes and orientifold planes.
Due to the boundary at the real axis there are non--trivial correlators between
left-- and right--movers. In order to compute the amplitudes, it is convenient to use
the ``doubling trick,'' \refs{\GubserWT,\GarousiAD} to convert disk correlators to the standard holomorphic ones.
This method  accommodates the boundary conditions by extending the definition of holomorphic fields to the entire complex plane
such that their operator product expansions (OPEs) on the complex plane reproduce all the OPEs among holomorphic and anti--holomorphic fields on $\IH$ \KosteleckyPX.

By the boundary conditions on the D--brane world--volume the open string momenta $p_i$ are restricted to lie within the world--volume directions. On the other hand, the closed string momentum $q$  has generic
directions. All strings are massless and their momenta are on-shell, i.e.  $p_i^2=q^2=0$.
Since D--branes are infinitely heavy objects they can absorb momentum in the transverse direction, which in turn implies that only along the world--volume directions momentum conservation is furnished. This can be taken into account by choosing
\eqn\Dbrane{
q_1=\h\ q\ \ \ ,\ \ \ q_2=\h\ Dq\ ,}
with $q$ the closed string momentum and $D$ a matrix accounting for the specific boundary conditions
in $d$ space--time dimensions.
Then, the longitudinal closed string momentum is given by
\eqn\totalclosed{
q^\parallel=q_1+q_2=\h\ (q+Dq)\ ,}
while normal to the brane we have the remaining momentum
\eqn\remaing{
q^\perp=\h\ (q-Dq)\ ,}
and total momentum conservation along the D--brane world volume reads:
\eqn\conservation{
\sum_{i=1}^{N-2}p_i+q^\parallel=0\ .}
Typically, in flat space--time\foot{The most general expression for $D$ is given  by $D=-g^{-1}+2\;(g+b)^{-1}$, with the metric $g$ and the anti--symmetric tensor $b$ \LustCX.} the matrix $D^{\mu\nu}$ is a diagonal matrix, equal to Minkowski metric $\eta^{\mu\nu}$ in directions along the D--brane
(Neumann boundary conditions) and to $-\eta^{\mu\nu}$ in directions orthogonal to the brane (Dirichlet boundary conditions). Then, the left-- and right--moving
momenta $q_i$ define on--shell momenta:
\eqn\onshell{
q_i^2=0\ .}
In what follows, we shall assume that \onshell\ holds\foot{Note, that
this assumption is obeyed by generic four--dimensional string compactifications with internal metric $g$ and two--form fluxes $b$ without warping for which a CFT description is available.} for the left-- and right--moving
closed string momenta \Dbrane.

The disk amplitudes involve integrals of the form
\eqn\GENERIC{\eqalign{
F_N&=V_{\rm CKG}^{-1}\ \delta\Big(\sum_{i=1}^{N-2}p_i+q_1+q_2\Big)\int\prod_{i=1}^{N-2} dx_i
\prod_{1\leq r<s\leq N-2}|x_r-x_s|^{2\alpha'p_rp_s}\ (x_r-x_s)^{n_{rs}}\cr
&\times \int_{\IH} d^2z\ (z-\ov z)^{2\al'q_1q_2+n} \ \prod_{i=1}^{N-2}\ (x_i-z)^{2\al'p_iq_1+n_i}\ (x_i-\ov z)^{2\al'p_iq_2+\ov n_i}\ ,}}
where we included the momentum--conserving (along the D-brane world--volume) \totalclosed\ delta function and divided by the volume $V_{\rm CKG}$ of the conformal Killing group.
The powers $n_{rs},~n_i,~\ov n_i,~ n$ are some integer numbers. To be specific, we focus on the amplitude associated to
one particular Chan-Paton factor (partial amplitude), ${\rm Tr}(T^1T^2\dots T^{N-2})$, with the integral over ordered  $x_1<x_2<\dots<x_{N{-}2}$. Note, that in \GENERIC, the momenta
$q_1$ and $q_2$ are assumed to be unrelated, i.e. in \Dbrane\ the matrix $D$ is a generic matrix such that
the condition \onshell\ is fulfilled.

For the concrete case \GENERIC, we write the integral over the complex upper half--plane $\IH$ as an integral
over holomorphic and anti--holomorphic coordinates, by following the method proposed in \KawaiXQ.
After writing $z=z_{1}+i z_{2}$, the integrand becomes an analytic function of $z_{2}$ with
$2(N{-}2)$ branch points at $\pm i(x_i-z_1)$.
We then deform the $z_{2}$--integral along the real axis $\im(z_{2})=0$ to the
pure imaginary axis $\re(z_{2})=0$ with $\im(z_{2})\geq 0$, as depicted in Fig. 1.
\bigskip
\centerline{\epsfxsize=0.7\hsize\epsfbox{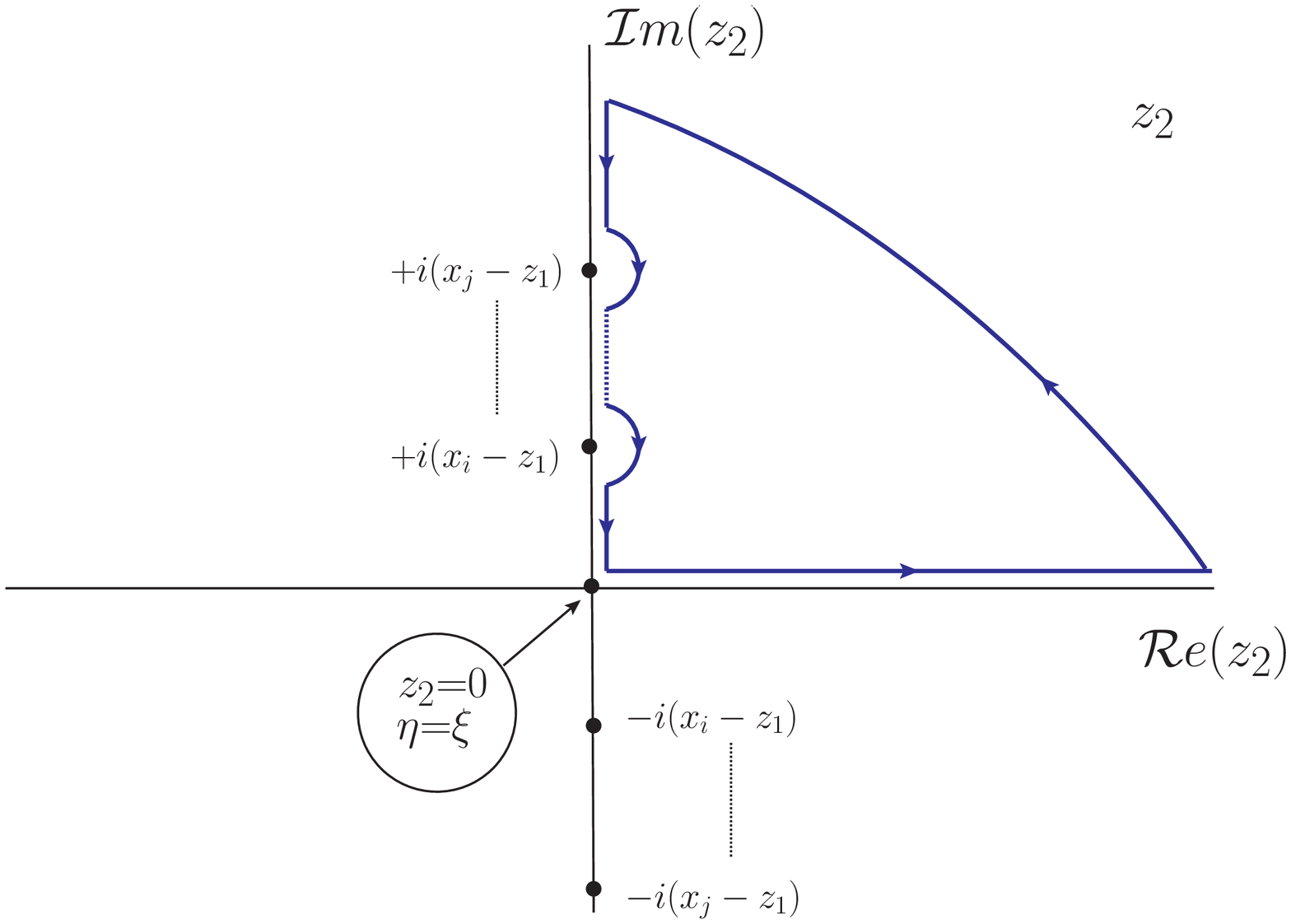}}
\noindent{\ninepoint\sl \baselineskip=8pt \centerline{{\bf Figure 1}: \sl
Branch cut structure and contour deformation in complex $z_2$--plane.}}\bigskip\bigskip
\noindent
In this way, the variables
\eqn\real{
\xi=z_{1}+i\ z_{2}\equiv z\ \ \ ,\ \ \ \eta=z_{1}-i\ z_{2}\equiv \ov z}
become real, subject to:
\eqn\subject{
\eta-\xi>0\ .}
After changing the integration variables $(z_1,z_2)\to (\xi,\eta)$ (with the Jacobian  $\det\fc{\p(z_{1},z_{2})}{\p(\xi,\eta)}=\fc{i}{2}$), Eq. \GENERIC\ becomes an integral over $N$ real positions $x_i,\xi,\eta$
\eqn\AMPLITUDE{\eqalign{
F_N=V_{\rm CKG}^{-1}\ &\delta\Big(\sum_{i=1}^{N}k_i\Big)\int\prod_{i=1}^{N-2} dx_i
\int_{-\infty}^\infty d\xi\int_{\xi}^\infty d\eta\!
\prod_{1\leq r<s\leq N-2}|x_r-x_s|^{2\alpha'k_rk_s}\ (x_r-x_s)^{n_{rs}}\cr
&\times\fc{i}{2}\  (\xi-\eta)^n\ |\xi-\eta|^{2\alpha'k_{N-1}k_N}\ \Om(\xi,\eta)\cr
&\times\prod_{i=1}^{N-2}\Pi(x_i,\xi,\eta)\ |x_i-\xi|^{2\alpha'k_ik_{N{-}1}}\ |x_i-\eta|^{2\alpha'k_ik_N}(x_i-\xi)^{n_i}(x_i-\eta)^{\bar n_i}\ ,}}
with the open string momenta $k_r=p_r,\ r=1,\ldots,N{-}2$ and the closed string momentum split into  left-- and right--moving parts
\eqn\split{
k_{N{-}1}=q_1\ \ \ ,\ \ \ k_N=q_2\ ,}
respectively.
Eq. \AMPLITUDE\ resembles a generic  open string integral involving $N$ open strings with external
momenta $k_i$ supplemented by the  extra phase factors
\eqn\PHASE{\eqalign{
\Om(\xi,\eta)&=e^{2\pi i\al'k_{N-1}k_N \,\theta(\eta-\xi)}\ ,\cr
\Pi(x_i,\xi,\eta)&=e^{-2\pi i\al'k_ik_{N-1} \, \theta(\xi-x_i)}\ e^{2\pi i\al'k_ik_{N} \, \theta(\eta-x_i)} \ ,}}
where $\theta$ denotes the Heaviside step function.
These monodromy factors \PHASE\ account for the
correct branch of the integrand,
making the integral well defined.
 Note that the phases, which are independent on the integers
$n_{rs},n_i,\ov n_i, n$ do not depend on the particular values of  integration variables, but only on the ordering of $\xi$ and $\eta$ with respect to the original $N{-}2$ vertex positions.
In this way, the original integral becomes a weighted (by phase factors) sum of integrals, each of them having the same form as the integrals appearing in $N$-point (partial) open string amplitudes, with the vertices inserted at  $x_l, ~l=1,\dots,N$, where we identified $x_{N{-}1}\equiv\xi$  and $x_N\equiv\eta$.
Note that the order of the original $N{-}2$ positions remains unchanged. Since the closed string vertex factorizes into two gauge bosons inserted at $z=\xi=x_{N{-}1}$ and  $\ov z=\eta=x_N $, we conclude that
the amplitude
\eqn\trivial{
F_N\equiv A(1,2,\dots,N{-}2;q_1,q_2)}
describing closed string decay into  $N{-}2$ gauge bosons can be written as a weighted sum of pure open string amplitudes with the closed string replaced by a pair of gauge bosons. The latter carry the left-- and right--moving momenta Eq. \split\ of the closed string, respectively.

In order to express the partial amplitude $A(1,2,\dots,N{-}2;q_1,q_2)$ in terms of $N$-point open string amplitudes, we need to analyze the phase factors.
For $\xi\in\IR$ the phase factor \PHASE\ in the integrand  can be accommodated by considering respective contours in the complex  $\eta$--plane.
After fixing the position of the first open string vertex at $x_1=-\infty$ we have the situation depicted in Fig. 2.
For the case of interest $\eta>\xi$ quite generally, around all open string vertex positions  $x_l>\xi$  the contour goes  anti--clockwise.
The last case  $\xi>x_{N-2}$ contributes the single term
\eqn\contrib{
\fc{i}{2}\ \exp\lf\{\pi i \lf(s_{1,N-1}+s_{N-1,N}-s_{1,N}\ri)\ri\}\,A(1,2,\ldots,N-2,N-1,N)\ ,}
while the first case $\xi<x_2$ gives rise to:
\eqn\contria{
-\fc{i}{2}\ A(1,N,N-1,2,\ldots,N-2)\ .}

\medskip
\centerline{\epsfxsize=0.8\hsize\epsfbox{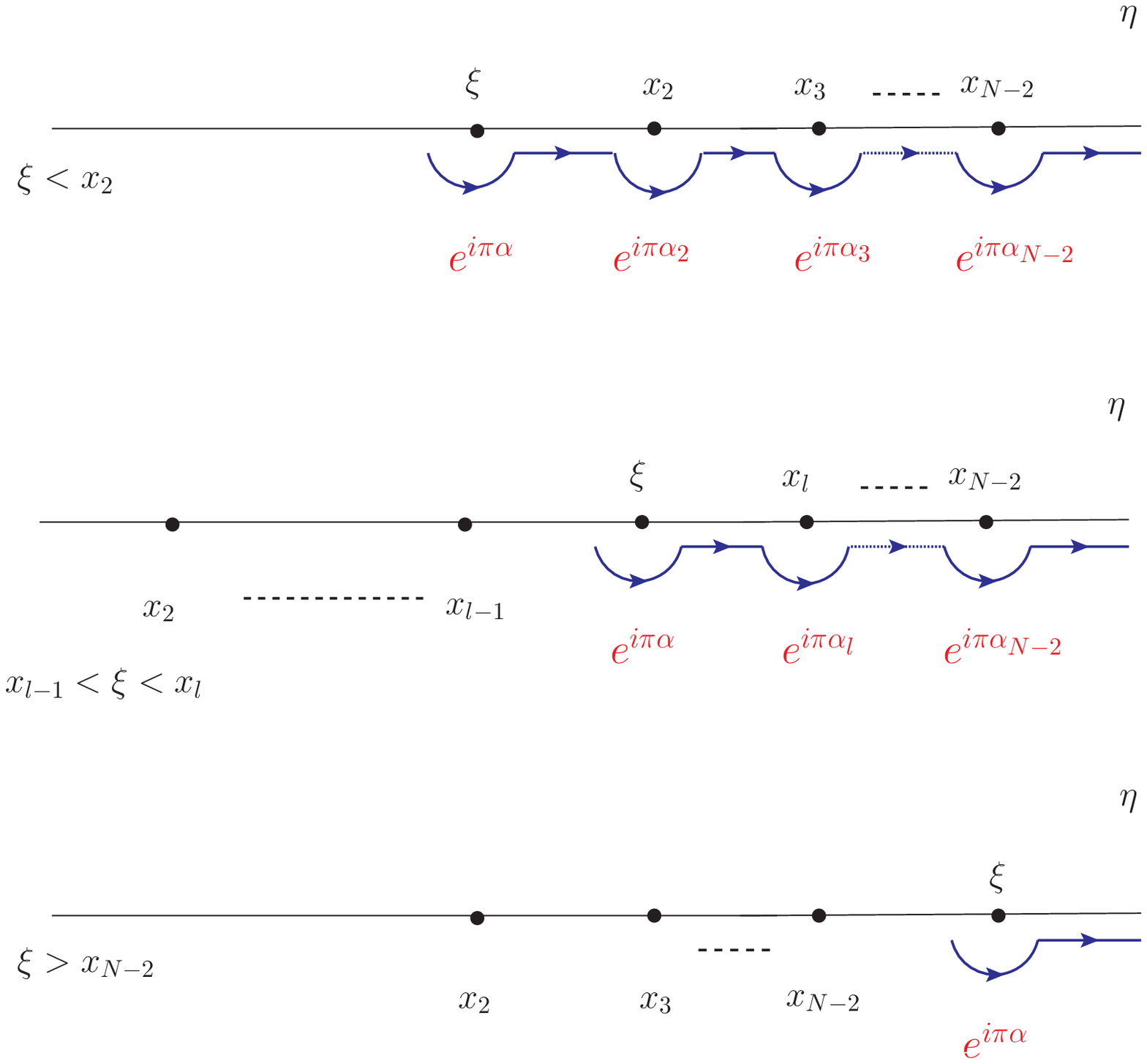}}
\noindent{\ninepoint\sl \baselineskip=8pt \centerline{{\bf Figure 2}: \sl
Complex $\eta$--plane and contour integrations.}
\centerline{\sl Here $\alpha_l\equiv\alpha'p_lq_2=2\alpha'k_lk_N$ and $\alpha\equiv2\ap q_1q_2=2\ap k_{N-1}k_N$.}}
\bigskip\bigskip
\noindent
In the latter case we could reduce the contour  to a single contribution
thanks to string monodromy relations \StiebergerHQ.
Eventually, in the second case $x_{l-1}<\xi<x_l$ with $l=2,\ldots,N-2$
string monodromy relations can be applied to deform the contour to the left.
This is accomplished for $x_2<\xi<x_{\ceiling{\fc{N}{2}}}$ to obtain a minimal set of integration regions. Each case  $x_l<\xi<x_{l+1}$ with $l=2,\ldots,\ceiling{\fc{N}{2}}-1$ contributes
 a residual contour of $l$ arcs starting from $x_1=-\infty$ and passing the $l$
points  $x_2,\ldots,x_l$ and $\xi$:
\eqnn\contric{
$$\eqalignno{
-\fc{i}{2}\ &\sum_{l=2}^{\ceiling{\fc{N}{2}}-1}\ \sum_{i=2}^{l+1}
\exp\lf\{\pi i \lf(\sum_{j=2}^{i-1}s_{j,N}-\sum_{j=2}^ls_{j,N-1}\ri)\ri\}\cr \crr
&\times  A(1,\ldots,i-1,N,i,\dots,l,N-1,l+1,\ldots,N-2)\ .&\contric}
$$}
On the other hand, for $x_{\ceiling{\fc{N}{2}}}<\xi<x_{N-2}$ we leave the contour as depicted in Fig. 2 and obtain  contributions from  each region $x_l<\xi<x_{l+1}$ with
$l=\ceiling{\fc{N}{2}},\ldots,N-3$. Each giving rise to a contour from $\xi$ to  infinity
with $N-1-l$ arcs at the points  $\xi$ and   $x_{l+1},\ldots, x_{N-2}$:
\eqnn\contrid{
$$\eqalignno{
+\fc{i}{2}\ &\sum_{l=\ceiling{\fc{N}{2}}}^{N-3}\ \sum_{i=l}^{N-2}
\exp\lf\{\pi i \lf(s_{1,N-1}+s_{N-1,N}-s_{1,N}+\sum_{j=l+1}^{N-2}s_{j,N-1}-
\sum_{j=i+1}^{N-2}s_{j,N}\ri)\ri\}\cr \crr
&\times A(1,\ldots,l,N-1,l+1,\ldots,i,N,i+1,\ldots,N-2)\ .&\contrid}
$$}
In total we obtain $\floor{\fc{N}{2}}\ (\ceiling{\fc{N}{2}}-1)$ terms:
\eqnn\TOTAL{
$$\eqalignno{
A(1,\ldots,&N-2;q_1,q_2)=-\fc{i}{2}\ \sum_{l=1}^{\ceiling{\fc{N}{2}}-1}\ \sum_{i=2}^{l+1}
\exp\lf\{\pi i \lf(\sum_{j=2}^{i-1}s_{j,N}-\sum_{j=2}^ls_{j,N-1}\ri)\ri\}\cr\crr
&\times A(1,\ldots,i-1,N,i,\dots,l,N-1,l+1,\ldots,N-2)\cr\crr
&+\fc{i}{2}\ \sum_{l=\ceiling{\fc{N}{2}}}^{N-2}\ \sum_{i=l}^{N-2}
\exp\lf\{\pi i \lf(s_{1,N-1}+s_{N-1,N}-s_{1,N}+\sum_{j=l+1}^{N-2}s_{j,N-1}-
\sum_{j=i+1}^{N-2}s_{j,N}\ri)\ri\}\cr\crr
&\times A(1,\ldots,l,N-1,l+1,\ldots,i,N,i+1,\ldots,N-2)\ .&\TOTAL}
$$}
On the r.h.s., according to \split\ the $N$ open string momenta are given by
$k_i,\ i=1,\ldots,N-2$, $k_{N-1}=q_1$ and $k_N=q_2$, respectively.
Furthermore, with \Dbrane\ and \totalclosed\ we may express the kinematic invariants
\eqn\INV{\eqalign{
s_{N-1,N}&=\ap\; (q^\parallel)^2\ ,\cr
s_{i,N-1}&=\ap\; p_iq^\parallel\ ,\ s_{i,N}=\ap\; p_iDq^\parallel\ \ \ ,\ \ \ i=1,\ldots,N-2\ ,}}
in terms of invariants of the D--brane world--volume (i.e. using only momenta parallel to the D--brane world--volume)
with the parallel closed string momentum $q^\parallel$ defined in \totalclosed.
It is easy to see, that for $q_1=q_2$, \ie $s_{1,N-1}=s_{1,N}$ and
$s_{N-1,N}=0$ the real part of \TOTAL\ reduces to the formula given in \StiebergerCEA\
describing the result in the double cover.

As an example we display the case $N=5$ for which \eqq \TOTAL\ yields  the following four terms:
\eqnn\TOTALfour{
$$\eqalignno{
A(1,2,3;q_1,q_2)&=\fc{i}{2}\ \Big\{\ e^{\pi i (s_{14}-s_{15}+s_{45})}\ A(1,2,3,4,5)-
e^{\pi i (-s_{24}+s_{25})}\ A(1,2,5,4,3)\cr
&-e^{-\pi i s_{24}} A(1,5,2,4,3)-A(1,5,4,2,3)\ \Big\}\ .&\TOTALfour}$$}

By applying string monodromy relations \refs{\StiebergerHQ,\BjerrumBohrRD}  the expression \TOTAL\ can be expressed in terms of the minimal set of $(N-3)!$ open string basis amplitudes:
\eqnn\FINAL{
$$\eqalignno{A(1,2,\dots,N{-}2;q_1,q_2)&=(-1)^N\ e^{-\pi i (s_{1,N}+s_{2,N-1})}\
\sum_{l=2}^{N-2}(-1)^l\ \sin(\pi s_{l,N-1})\ e^{\pi i (-1)^l\ s_{l,N-1}} \cr
&\times\sum_{\rho\in \{OP(\al,\bet^t),l\}}
e^{\pi i\sum\limits_{k=1}^{\floor{\fc{N-3}{2}}}\tau_{2k+1}(\rho)}\
\Sc(\rho)\ A(1,\rho,N-1,N)\ .&\FINAL}$$}
The second sum involves all  permutations $\rho$ comprising the element $l$ and the ordered set of permutations $OP(\al,\bet^t)$ of the merged sets:
\eqn\sets{
\al=\{2,\ldots,l-1\}\ \ \ ,\ \ \ \bet=\{l+1,\ldots,N-2\}\ .}
This ordered set corresponds to all permutations of $\al\cup \bet^t$ which keep the order of elements
of $\al$ and $\bet^t$, respectively. Besides, $\bet^t$ denotes reversal of the elements in $\bet$.
Furthermore, in \FINAL\ the following string kernel $\Sc(\rho)$ enters
\eqn\kernel{
\Sc(\rho)\equiv \Sc[ \rho(2,\ldots,N-2) \, ] = \prod_{i=2}^{N-2}\prod_{j=i+1}^{N-2}
\exp\lf\{  \pi i\;\Theta(\rho^{-1}(i)-\rho^{-1}(j)) \ s_{i,j} \ri\}\ ,}
with $s_{i,j}\equiv s_{ij}= 2\alpha'k_ik_j$ and the Heaviside step function
$\Theta$. Some other variants of string KLT kernels occur for pure closed string amplitudes in \refs{\KawaiXQ,\Bohr}. Finally, we have:
\eqn\TAU{
\tau_i(\rho)=\cases{
{\rm sign}(\rho^{-1}(i)-\rho^{-1}(i+1))\ (s_{i,N-1}+s_{i+1,N-1})\ ,&$3\leq i\leq N-3$\ ,\cr
s_{N-2,N-1}\ ,&$i=N\!-\!2$\ .}}
In \FINAL\ the double sum gives rise to $\sum\limits_{l=2}^{N-2}\lf(N-4\atop l-2\ri)=2^{N-4}$ terms.
It can be evidenced along the lines of \refs{\StiebergerTE,\MafraNVii}, that  \FINAL\ provides the correct soft--limits for $k_{N-2}\ra 0$ and collinear limits.

Note, that the leading term in the $\ap$--expansion of \FINAL\ starts linearly at $\ap$, while
the leading term in \TOTAL\ appears at $\ap^0$. As a consequence, the latter must vanish, giving rise to relations similar to (and following from) $U(1)$ decoupling (Kleiss-Kuijf) \KleissNE\ conditions,
\eqnn\KK{
$$\eqalignno{
0&=-\sum_{l=1}^{\ceiling{\fc{N}{2}}-1}\ \sum_{i=2}^{l+1}
 A_{YM}(1,\ldots,i-1,N,i,\dots,l,N-1,l+1,\ldots,N-2)\cr\crr
&+\sum_{l=\ceiling{\fc{N}{2}}}^{N-2}\ \sum_{i=l}^{N-2}
 A_{YM}(1,\ldots,l,N-1,l+1,\ldots,i,N,i+1,\ldots,N-2)\ ,&\KK}
$$}
involving $\floor{\fc{N}{2}}\ (\ceiling{\fc{N}{2}}-1)$ SYM subamplitudes.

To illustrate the result \FINAL\ let us consider some examples with a small number of external particles.
The case $N=4$ is not new and has already been studied in \refs{\HashimotoKF,\GarousiAD}.
For completeness we display the latter and \FINAL\ yields:
\eqn\finalff{
A(1,2;q_1,q_2)=e^{-\pi i s_{23}}\ \sin(\pi s_{23})\ A(1,2,3,4)\ .}
In the appendix we explicitly demonstrate how to cast a generic mixed amplitude of two open and one closed string into the form \finalff.
The real part gives the corresponding relation in the  double cover
(in this case we have $u=s_{23}=s_{24}$ and $s=s_{12}=-2s_{24}$):
\eqn\Refinalff{\eqalign{
A(1,2;q,q)&=\h\cos(\pi u)\ \sin(\pi u)\ A(1,2,3,4)=\sin(2\pi u)\ A(1,2,3,4)\cr
&=-\sin(\pi s)\ A(1,2,3,4)\ .}}
For $N=5$ our formula \FINAL\ yields
\eqn\finalf{
A(1,2,3;q_1,q_2)=e^{-\pi i s_{24}}\ \lf[\ e^{-\pi i s_{51}}\ \sin(\pi s_{34})\ A(1,2,3,4,5)-
\sin(\pi s_{24})\ A(1,3,2,4,5)\ \ri]\ ,}
which agrees with \TOTALfour. Again, in the appendix we explicitly demonstrate how to cast a generic mixed amplitude of three open and one closed string into the form \finalf.
The real part of \finalf\ gives the corresponding relation in the  double cover. After using open string
relations we obtain (in this case we have $s_{12}=2s+\al,\ s_{23}=2u+\al,\ s_{34}=s,\ s_{45}=\al,\ s_{51}=u$, i.e.
$\al_1=\al_2=s$ and $\bet_1=\bet_2=-s-u-\al=t$):
\eqn\finalfa{
A(1,2,3;q,q)=-\h\ \sin(\pi \al)\ A(1,2,3,4,5)-\h\ \sin(\pi t)\ A(1,5,2,4,3)\ ,}
in agreement with Eq. (3.19) of \StiebergerHQ.
For $N=6$ we find:
\eqn\six{\eqalign{
A(1,2,3,4;q_1,q_2)&=e^{-\pi i s_{25}}\ \lf\{\ e^{-\pi i (s_{61}+s_{35})}\ \sin(\pi s_{45})\ A(1,2,3,4,5,6)\ri.\cr
&+\sin(\pi s_{25})\ A(1,4,3,2,5,6)-e^{\pi i (-s_{61}+s_{34}+s_{45})}\ \sin(\pi s_{35})\cr
&\lf.\times\lf[\ A(1,2,4,3,5,6)+e^{\pi i s_{24}}\ \sin(\pi s_{35})\ A(1,4,2,3,5,6)\ \ri]\ \ri\}.}}
{}For $N=7$ we obtain:
\eqnn\seven{
$$\eqalignno{
A(1,2,3,4,5;q_1,q_2)&=e^{-\pi i s_{26}}\ \lf\{\ e^{-\pi i (s_{17}+s_{36}+s_{46})}\ \sin(\pi s_{56})\ A(1,2,3,4,5,6,7)\ri.\cr
&- \sin(\pi s_{26})\ A(1,5,4,3,2,6,7)-e^{-\pi i (s_{71}+s_{36}-s_{45})}\ \sin(\pi s_{46})\cr
&\times\lf[\ e^{\pi i s_{56}}\   A(1,2,3,5,4,6,7)-e^{\pi i (s_{56}+s_{35})}\  A(1,2,5,3,4,6,7)\ri.\cr
&\lf.-e^{\pi i (s_{56}+s_{25}+s_{35} )}\   A(1,5,2,3,4,6,7)\ \ri]+e^{-\pi i s_{36}}\ \sin(\pi s_{36})\cr
&\times\lf[\ e^{\pi i (s_{12}+s_{27})}\ A(1,2,5,4,3,6,7) +e^{-\pi i (s_{23}+s_{24}+s_{26})}\    A(1,5,2,4,3,6,7)\ri.\cr
&\lf.\lf.+e^{-\pi i (s_{23}+s_{26})}\  A(1,5,4,2,3,6,7)\ \ri]\ri\}\ .&\seven}$$}

Note that until this point, we did not make any assumption how the total closed string momentum $q$ was distributed
among left-- and right--movers. In particular, we did not use any specific form of the $D$ matrix, see Eq. \Dbrane.
We used the on--shell condition \onshell\ and the total momentum conservation \conservation\ only.
This should be contrasted with the computations on the disk double cover which utilize left--right
symmetric (half-half) momentum distribution. In order to make contact
with the results of \StiebergerKIA, we consider the case of a  D-brane filling four spacetime dimensions and the closed string carrying a purely four--dimensional momentum
\eqn\ppp{P=q=q_1+q_2~,}
which is on--shell, $P^2=0$, and now split as:
\eqn\split{
q_1= k_{N-1}=x\ P\ \ \ ,\ \ \ q_2=k_N=(1-x)\ P\ .}
Hence all results from before can be used for this case. For the invariants
\INV, we have
\eqn\INVV{\eqalign{
s_{N-1,N}&=0\ ,\cr
s_{i,N-1}&= x\, s_{iP} ,\quad s_{i,N}= (1-x)\, s_{iP} \ \ ,\qquad i=1,\ldots,N-2\ ,}}
where $s_{iP}=\ap p_iP$. We are interested in the field theory limit of the amplitudes, i.e.\ in the Einstein-Yang-Mills (EYM) limit which corresponds to the leading $\ap$ order of Eqs. \TOTAL\ and \FINAL. As mentioned before, at the $\ap^0$ order, the r.h.s.\ of Eq. \TOTAL\ vanishes as a result of \KK. At the leading $\ap$ order, Eq. \TOTAL\ yields:
\eqnn\FINALL{
$$\eqalignno{\quad & A_{\rm EYM}(1,2,\dots,N{-}2;P)=&\FINALL\cr
\ &\ \ {\pi\over 2}\ x\ \Biggl\{ \!\sum_{l=2}^{\ceiling{\fc{N}{2}}-1}\sum_{i=2}^l
\Big(\sum_{j=i}^ls_{jP}\Big)\ A_{\rm YM}(1,\ldots,i-1,N,i,\dots,l,N{-}1,l+1,\ldots,N{-}2)\cr
\,\,+&\sum_{l=\ceiling{\fc{N}{2}}}^{N-3}\sum_{i=l+1}^{N-2}
\Big(\!\sum_{j=l+1}^is_{jP}\Big)\ A_{\rm YM}(1,\ldots,l,N{-}1,l+1,\ldots,i,N,i+1,\ldots,N{-}2)\Biggr\}\cr
\ & +{\pi\over 2}\,(2x-1)\ \Biggl\{ \!\sum_{l=2}^{\ceiling{\fc{N}{2}}-1}\sum_{i=2}^{l+1}
\Big(\sum_{j=2}^{i-1}s_{jP}\Big)\ A_{\rm YM}(1,\ldots,i-1,N,i,\dots,l,N{-}1,l+1,\ldots,N{-}2)\cr
\,\,-&\sum_{l=\ceiling{\fc{N}{2}}}^{N-2}\sum_{i=l}^{N-2}
\Big(\!\sum_{j=2}^is_{jP}\Big)\ A_{\rm YM}(1,\ldots,l,N{-}1,l+1,\ldots,i,N,i+1,\ldots,N{-}2)\Biggr\}
\ .}
$$}
By a repeated use of Bern-Carrasco-Johansson \BernQJ\ and Kleiss-Kuijf \KleissNE\ relations, one can show that the terms enclosed by the second curly bracket are equal to $(-x)$ times the terms enclosed by the first bracket. In this way, we obtain:
\eqnn\FINN{
$$\eqalignno{\quad  A_{\rm EYM}&(1,2,\dots,N{-}2;P)=&\FINALL\cr
\! \!\!&\!\!\! {\pi}\, x(1-x)\ \Biggl\{ \!\sum_{l=2}^{\ceiling{\fc{N}{2}}-1}\sum_{i=2}^l
\Big(\sum_{j=i}^ls_{jP}\Big)\ A_{\rm YM}(1,\ldots,i-1,N,i,\dots,l,N{-}1,l+1,\ldots,N{-}2)\cr
\qquad+&\sum_{l=\ceiling{\fc{N}{2}}}^{N-3}\sum_{i=l+1}^{N-2}
\Big(\!\sum_{j=l+1}^is_{jP}\Big)\ A_{\rm YM}(1,\ldots,l,N{-}1,l+1,\ldots,i,N,i+1,\ldots,N{-}2)\Biggr\}
\ .}
$$}
The above result reproduces Eq. (8) of Ref. \StiebergerKIA, modulo the $\pi$ factor which together with $\ap$ combine into the gravitational coupling constant. Similarly, the leading $\ap$ order of
Eq. \FINAL\ has the same form as Eq. (18) of Ref. \StiebergerKIA. The set of permutations appearing in the sum is specified in Eq. \sets.

The advantage of the formalism developed in this work is that is that it allows for a left--right asymmetric partition of the closed string momentum. By considering the monodromy properties of the amplitude on the underlying world--sheet,
we derived Eq. \FINAL\ which shows that the full--fledged string disk amplitude involving one closed string and any number of open strings can be expressed as a linear combination of pure open string amplitudes with the original closed string momentum arbitrarily split between two open strings.

\vskip0.5cm
\goodbreak
\leftline{\noindent{\it Appendix}}

\br
\noindent
In this appendix we demonstrate how performing a direct complex world--sheet integration for the cases $N=4$ and $N=5$
readily leads to the results \finalff\ and \finalf, respectively.

Let us consider the world--sheet disk integral involving two open and one closed string
\eqnn\mixedff{
$$\eqalignno{
F_4&=\int_{-\infty}^{+\infty} dx \ (2i)^{\alpha_0}\ (x-i)^{\alpha_1}\ (x+i)^{\alpha_2}=-\pi\ e^{-\pi i \al_1}\ \fc{\Gamma(-1-\al_1-\al_2)}{\Gamma(-\al_1)\ \Gamma(-\al_2)}\cr
&= \sin(\pi\al_1)\ e^{-\pi i \al_1}\ \fc{\Gamma(1+\al_0)\ \Gamma(1+\al_1)}{\Gamma(-\al_2)}\ ,&\mixedff}$$}
corresponding to the choice of vertex positions:
\eqn\jix{
z_1=-\infty\ ,\ z_2=x,\ z_3=i,\ \ov z_3=-i\ .}
Above we have imposed the constraint:
\eqn\condff{
\al_0+\al_1+\al_2=-2\ .}
The integral \mixedff\ can be computed by considering the contours in the complex $x$--plane
as shown in Fig. 3.

\bigskip
\centerline{\epsfxsize=0.6\hsize\epsfbox{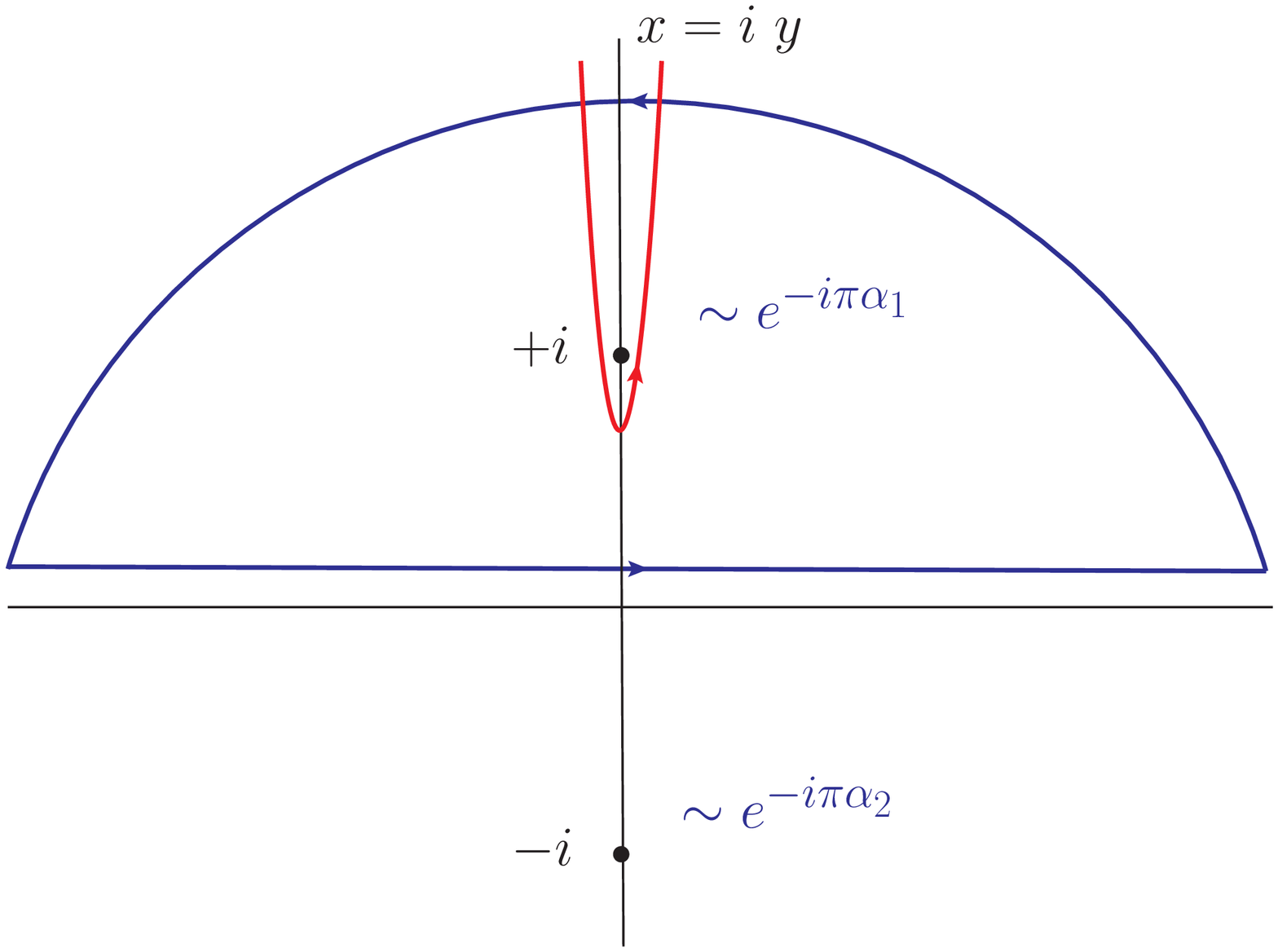}}
\noindent{\ninepoint\sl \baselineskip=8pt \centerline{{\bf Figure 3}: \sl
Contour integration in the complex $x$--plane.}}
\bigskip\bigskip
\noindent
After deforming the integration from the real axis to the imaginary axis we obtain
$$\eqalignno{
&(2i)^{\al_0}\ (e^{\pi i \alpha_1}-e^{-\pi i \alpha_1})\ \int_1^\infty dy\ (y-1)^{\alpha_1}\ (y+1)^{\alpha_2}\ i^{\al_1+\al_2+1}\cr
&=(2i)^{\al_1+\al_2+\al_0+2}\ \sin(\pi\al_1)\ \fc{\Gamma(1+\al_1)\ \Gamma(-1-\al_1-\al_2)}{\Gamma(-\al_2)}\ ,}
$$
which yields\foot{Later we will use the integral:
\eqn\intff{
\int_{-\infty}^{+\infty} dx\ (x-i)^a\ (x+i)^b=-\pi\ (2i)^{2+a+b}\ e^{-\pi i a}\
\fc{\Gamma(-1-a-b)}{\Gamma(-a)\ \Gamma(-b)}\ .}}
 \mixedff\ subject to the branching factor $e^{-\pi i\al_1}$.
We can relate the mixed amplitude \mixedff\ to a pure open string amplitude involving four open strings.
The generic open string disk amplitude reads:
\eqn\FOURSTR{
A(1,2_\pi,3_\pi,4_\pi)=V_{\rm CKG}^{-1}\int_{D(\pi)} dz_i\ \prod_{i<j}^4|z_{ij}|^{s_{ij}}\ z_{ij}^{n_{ij}}\ .}
Note, that due to conformal invariance the integers $n_{ij}$ must fulfil the conditions:
\eqn\conditions{
\sum_{j\neq i}^N n_{ij}=-2\ ,\ i=1,\ldots,4\ .}
With the choice
\eqn\choc{
z_1=-\infty,\ z_2=0,\ z_3=1,\ z_4=x^{-1}\ ,}
the canonical subamplitude becomes
\eqnn\Openff{
$$\eqalignno{
A(1,2,3,4)&=(-1)^{n_{23}+n_{24}+n_{34}}\ \fc{\Gamma(-1-\tilde s_{24}-\tilde s_{34})\ \Gamma(1+\tilde s_{34})}{\Gamma(-\tilde s_{24})}\cr
&=(-1)^{n_{23}+n_{24}+n_{34}}\
\fc{\Gamma(1+\al_0)\ \Gamma(-1-\al_0-\al_2)}{\Gamma(-\al_2)}\ ,&\Openff}$$}
with
\eqn\witz{
\tilde s_{ij}=s_{ij}+n_{ij}\ ,}
and  the following identifications
\eqn\condsi{\eqalign{
\al_0&=s_{34}+n_{34}\ ,\cr
\al_1&=s_{23}+n_{23}\ ,\cr
\al_2&=s_{24}+n_{24}\ ,}}
which fulfil  \condff, \ie $n_{23}+n_{24}+n_{34}=-2$.
Comparing \mixedff\ with \Openff\ gives the relation \finalff\ subject to \trivial.

Now, let us consider the world--sheet disk integral involving three open and one closed string\foot{
This specific integral  has already been computed in \GarousiEA\ without making reference to
the underlying five--point open string amplitude. This link will be established in the sequel.}
\eqnn\mixedf{
$$\eqalignno{
F_5&=(2i)^{\alpha_0}(-1)^{n_{23}}\int_{-\infty}^{+\infty} dx_2 \int_{x_2}^\infty dx_3\ (x_2-i)^{\bet_2}\ (x_2+i)^{\bet_1}\ (x_3-i)^{\al_2}\ (x_3+i)^{\al_1}\cr
&\times(x_3-x_2)^{\bet_0}=-\ (-1)^{n_{23}}\ e^{-\pi i (\al_2+\bet_2)}\ \Gamma(-2-\al_1-\al_2-\bet_1-\bet_2-\bet_0)\cr
&\times\Big\{ \sin[\pi(\bet_0+\al_1+\bet_1)]\ \fc{\Gamma(1+\bet_0)\ \Gamma(-1-\al_1-\bet_0)\ \Gamma(2+\bet_0+\al_1+\bet_1)}{\Gamma(-\al_1)\ \Gamma(-\bet_2-\al_2)}\cr
&\times{}_3F_2\lf[{-\al_2,1+\bet_0,2+\bet_0+\al_1+\bet_1\atop 2+\bet_0+\al_1,-\al_2-\bet_2};1\ri]\cr
&+e^{-\pi i(\al_1+\bet_0)}\ \sin(\pi \bet_1)\ \fc{\Gamma(1+\bet_1)\ \Gamma(-1-\al_1-\al_2-\bet_0)\ \Gamma(1+\bet_0+\al_1)}{\Gamma(-\al_2)\ \Gamma(-1-\bet_0-\bet_2-\al_1-\al_2)}\cr
&\times{}_3F_2\lf[{-\al_1,-1-\al_1-\al_2-\bet_0,1+\bet_1\atop -\al_1-\bet_0,-1-\bet_0-\bet_2-\al_1-\al_2};1\ri]\Big\}\ ,&\mixedf}$$}
corresponding\foot{The factor $(-1)^{n_{23}}$ becomes obvious in the following.}
to the choice of vertex positions:
\eqn\FIXC{
z_1=-\infty\ ,\ z_2=x_2,\ z_3=x_3,\ z_4=i,\ \ov z_4=-i\ .}
Above we have used   \intff\ and the following integral:
$$\eqalignno{
\int_{0}^{+\infty} dx\ (x+\bet)^c\ (x+\gamma)^d\ x^e&=\bet^c\ \gamma^{1+e+d}
\ \fc{\Gamma(1+e)\ \Gamma(-1-c-d-e)}{\Gamma(-c-d)} \cr
&\times{}_2F_1\lf[{-c,1+e\atop -c-d};1-\fc{\gamma}{\beta}\ri]\ .}
$$
In addition, we have imposed the constraint:
\eqn\condf{
\al_0+\al_1+\al_2+\bet_0+\bet_1+\bet_2=-3\ .}
We can relate the mixed amplitude \mixedf\ to a pure open string amplitude involving five open strings.
The generic open string disk amplitude reads:
\eqn\FIVE{
A(1,2_\pi,3_\pi,4_\pi,5_\pi)=V_{\rm CKG}^{-1}\int_{D(\pi)} dz_i\ \prod_{i<j}^5 |z_{ij}|^{s_{ij}}\ z_{ij}^{n_{ij}}\ .}
Note, that due to conformal invariance the integers $n_{ij}$ must fulfil the conditions:
\eqn\conditions{
\sum_{j\neq i}^N n_{ij}=-2\ ,\ i=1,\ldots,5\ .}
The choice
\eqn\FIX{
z_1=-\infty,\ z_2=0,\ z_3=1,\ z_4=(xy)^{-1},\ z_5=x^{-1}}
gives rise to the subamplitude $A(1,2,3,5,4)$
\eqnn\Openfa{
$$\eqalignno{
A(1,2,3,5,4)&=(-1)^{n_{23}+n_{24}+n_{25}+n_{34}+n_{35}}\ \fc{\Gamma(1+\bet_0)\ \Gamma(1+\al_1)\ \Gamma(2+\al_1+\bet_1+\bet_0)\ \Gamma(1+\al_0)}{\Gamma(2+\bet_0+\al_1)\ \Gamma(-\al_2-\bet_2)}\cr
&\times{}_3F_{2}\lf[{-\al_2,1+\bet_0,2+\al_1+\bet_1+\bet_0\atop 2+\bet_0+\al_1,-\al_2-\bet_2}\ri]\ ,&\Openfa}
$$}
with  the following identifications:
\eqn\parameter{\eqalign{
\al_0&=s_{45}+n_{45}\ ,\ \bet_0=s_{23}+n_{23}\ ,\cr
\al_1&=s_{35}+n_{35}\ ,\ \bet_1=s_{25}+n_{25}\ ,\cr
\al_2&=s_{34}+n_{34}\ ,\ \bet_2=s_{24}+n_{24}\ ,}}
which fullfils  \condf, iff $n_{23}+n_{24}+n_{25}+n_{34}+n_{35}+n_{45}=-3$.
On the other hand, the choice
\eqn\fic{
z_1=-\infty,\ z_2=0,\ z_3=(xy)^{-1},\ \ z_4=x^{-1},\ z_5=1}
gives rise to the subamplitude $A(1,2,5,4,3)$:
\eqnn\Openfb{
$$\eqalignno{
A(1,2,5,4,3)&=(-1)^{n_{23}+n_{24}+n_{25}}\ \fc{\Gamma(1+\al_0)\ \Gamma(1+\bet_1)\ \Gamma(-1-\bet_0-\al_1-\al_2)\ \Gamma(1+\al_2)}{\Gamma(-1-\bet_0-\bet_2-\al_1-\al_2)\ \Gamma(-\al_1-\bet_0)}\cr
&\times{}_3F_{2}\lf[{-\al_1,1+\bet_1,-1-\al_1-\al_2-\bet_0\atop -\bet_0-\al_1,-1-\bet_0-\bet_2-\al_1-\al_2}\ri]\ .&\Openfb}
$$}
Comparing \mixedf\ with \Openfa\ and \Openfb\ gives the relation:
\eqnn\finalfa{
$$\eqalignno{
F_5&=e^{-\pi i(s_{24}+s_{34})}\ \Big\{\ \sin[\pi(s_{35}+s_{25}+s_{23})]\
\fc{\sin(\pi s_{35})}{\sin[\pi(s_{35}+s_{23})]}\ A(1,2,3,5,4)\cr
&-e^{-\pi i(s_{35}+s_{23})}\ \sin(\pi s_{25})\ \fc{\sin(\pi s_{34})}{\sin[\pi(s_{35}+s_{23})]}\
A(1,2,5,4,3)\ \Big\}\ .&\finalfa}
$$}
Eventually, after applying string monodromy relations involving five open strings \StiebergerHQ\ the expression \finalfa\ can be cast into \finalf\ subject to \trivial.

\vskip0.5cm
\goodbreak
\leftline{\noindent{\bf Acknowledgments}}

\noindent

This material is based in part upon work supported by the National Science Foundation under grant No.\  PHY-1314774.   Any
opinions, findings, and conclusions or recommendations expressed in
this material are those of the authors and do not necessarily reflect
the views of the National Science Foundation.

\listrefs

\end